# AI and conventional methods for UCT projection data estimation

Ankur Kumar[1], Prasunika Khare[1], Mayank Goswami[1#]

[1]Divyadrishti Laboratory, Department of Physics, IIT Roorkee, Roorkee, India

#mayank.goswami@ph.iitr.ac.in,

## Abstract

A 2D Compact ultrasound computerized tomography (UCT) system is developed. Fully automatic post processing tools involving signal and image processing are developed as well. Square of the amplitude values are used in transmission mode with natural 1.5 MHz frequency and rise time 10.4 ns and fall time 8.4 ns and duty cycle of 4.32%. Highest peak to corresponding trough values are considered as transmitting wave between transducers in direct line talk. Sensitivity analysis of methods to extract peak to corresponding trough per transducer are discussed in this paper. Total five methods are tested. These methods are taken from broad categories: (a) Conventional and (b) Artificial Intelligence (AI) based methods. Conventional methods, namely: (a) simple gradient based peak detection, (b) Fourier based, (c) wavelet transform are compared with AI based methods: (a) support vector machine (SVM), (b) artificial neural network (ANN). Classification step was performed as well to discard the signal which does not has contribution of transmission wave. It is found that conventional methods have better performance. Reconstruction error, accuracy, F-Score, recall, precision, specificity and MCC for 40 x 40 data 1600 data files are measured. Each data file contains 50,002 data point. Ten such data files are used for training the Neural Network. Each data file has 7/8 wave packets and each packet corresponds to one transmission amplitude data. Reconstruction error is found to be minimum for ANN method. Other performance indices show that FFT method is processing the UCT signal with best recovery.

## 1. Introduction

Computerized Tomography (CT) is an established non-destructive testing (NDT) technology to analyse the properties/inner profile of the specimen non-invasively. The X-ray, Ultrasound, Gamma-Ray, Positron, etc. are some means that are exploited for computerized tomography. Ultrasound is a preferred modality (for certain applications) used in the NDT due to the applicability, relatively cheap, and hazard-free operation [1–4]. The ultrasound based NDT systems involves signal processing techniques based on either (a) time of flight or (b) amplitude analysis [5]. The analysis of a specimen can be performed to inspect and detect the size and location of the defects. In time of flight analysis, the time taken by the signal to travel through the specimen contains the information about specimen's properties. While, in amplitude analysis is based on the amplitude of the received signal. The ultrasound CT (UCT) system involves emitter('s) and receiver('s) arranged in a specific manner (say fan beam or parallel beam), called transmission UCT. The emitter emits an ultrasound pulse of desired frequency which traverses through the specimen and finally detected by the detector. The collected data is processed to extract the projection data and used to reconstruct the profile of the specimen. This whole process requires a systematically collection of the data points for each scanning angle and translation (for each translation in case of single emitter-receiver UCT) and an efficient signal processing algorithm to automatically process all the data.

A typical UCT system operates at high frequency ranging from 100 kHz to 10 MHz. The high frequency signal requires a higher sampling rates to efficiently capture the signal which results in large data volumes [6]. This leads to time consuming signal processing sessions and cost. Usually, in an ultrasound signal there is large disparity in the useful and captured data points. This problem has been overcome by extracting only the time of flight data and/or amplitude analysis data[7]. For a transmission UCT system, acquisition of the time of flight data requires an additional circuitry and it is often contaminated by surrounding vibrations. The noise due to the measurement system, and interaction of ultrasound wave with the specimen also affects the signal. Generally, in many applications, the noise is assumed to be an uncorrelated Gaussian variable with approximately zero mean[8, 9]. Thus, the amplitude analysis based signal processing is more reliable for transmission mode UCT. The signal received by the receiver after passing through the specimen contains some non-linearity and distortion due to scattering, dispersion in addition to the inherited noise due to the electronics contribute to the error in extracting the projection data. The exact estimate of the signal amplitude can be extracted only if peaks and troughs are located in a synchronised manner i.e. peak and trough should have minimum time in forward manner.

In literature, several methods have been reported to detect the peaks for the different applications such as QRS complex detection in ECG signal, spectral peak detection. The conventional methods based on s-transform and Shannon energy, various thresholding methods, wavelet transform are mainly used to detect the peaks [10–14]. In ultrasound NDT systems, mainly time of flight method has been used to extract the useful information [7, 9, 15–17]. The artificial intelligence methods such as KNN (K- nearest neighbour), SVM (support vector machine), and ANN (artificial neural network) are used to process ultrasonic signal to detect the defects. The methods consists of pre-processing, feature extraction and neural network classification [18–20]. The amplitude analysis methods for extraction of projection data for a practical UCT application are not reported in literature to the best of our knowledge. In this work, we have developed signal processing methods for amplitude analysis based on the conventional and AI methods. An in-house developed 2D compact UCT system consisting of single emitter-receiver pair is used to generate ~~scan the specimen~~ data for testing and training.

## 1.1 Motivation:

A typical UCT system generates a large amount of data that increases manifold as we go for higher number of projections. Handling such a large data to extract projection data for further reconstruction is a cumbersome process. The distortion in the signal due to scattering, measurement system further increases the level of rigorousness. Thus, a suitable signal processing methodology needs to be applied. The method should collect the data automatically, and process it to extract the projection data. In this study, we have developed custom codes for signal processing based on amplitude analysis technique to automatically extract the projection data for the whole scanning process. The developed codes are based on the conventional signal processing techniques and artificial intelligence. The conventional method includes FFT transform, Wavelet transform and classical method. Artificial intelligence based methods includes KNN and SVM. The dataset to test the applicability of the methods is generated by in house developed automatic UCT system. A comparative study of these methods based on performance, time taken and applicability to a practical UCT system is carried out. Root mean square error (RMSE) is chosen as the performance parameter. The output of these methods have been applied for reconstruction of the specimen profile and included in the comparison.

## 2. Methodology

An in-house developed automatic 2D UCT system is used to scan the specimen as shown in figure 1. It comprises of the two non- contact focused ultrasound transducers tuneable between 100 kHz to 35 MHz frequency range, used as the emitter and receiver. Emitter is connected to an arbitrary wave generator and receiver to the data acquisition system via RG-316u BNC probes. The system does not require use of any couplant and medium to scan the specimen[21]. Thus, the system is straight forward to operate and free from any couplant

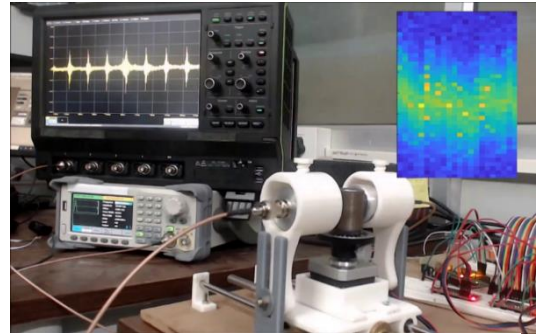

Figure 1 The UCT system used to generate data

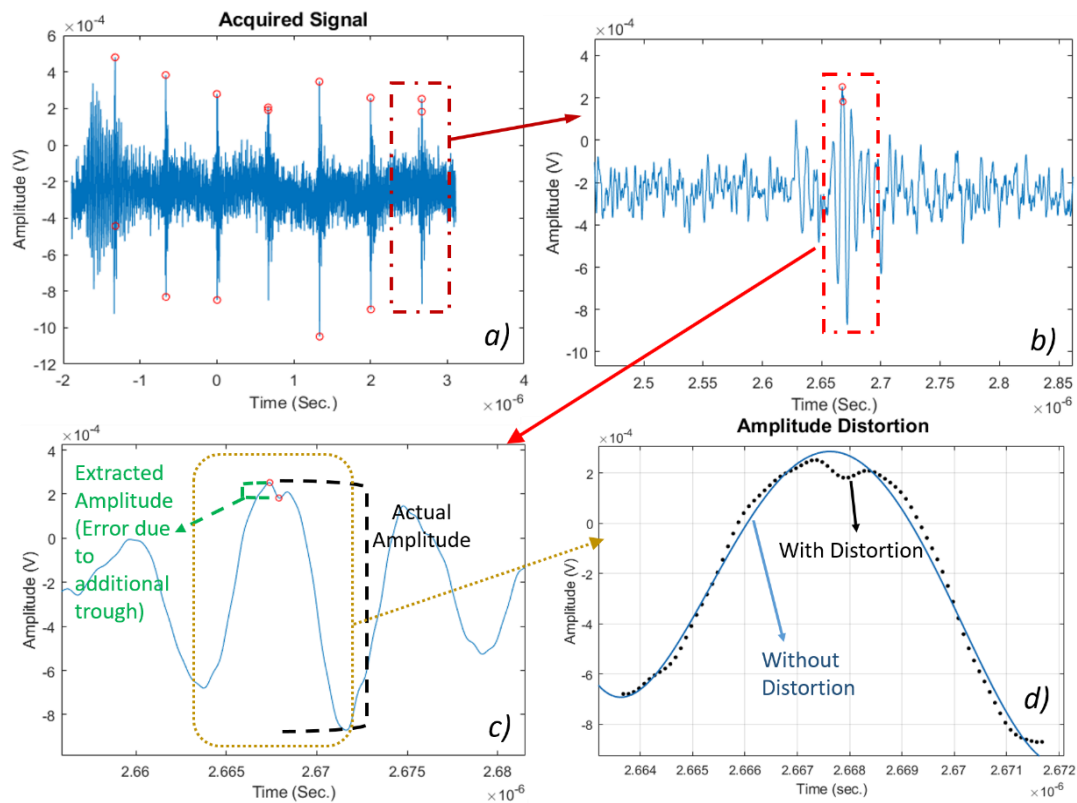

Figure 2 a) Acquired signal (7 wave packets), b) A single wave packet, c) Error in amplitude detection due to additional trough, d) Amplitude distortion.

related issues. The parallel beam profile is used to scan the specimen. A composite phantom of diameter 2.4 cm, made up of Steel (grade-80) and aluminium (grade-1050), with steel at outer periphery and aluminium at inner periphery is used as the specimen. The transducers position is kept fixed and the specimen is mounted on a controllable rotating platform. LabVIEW™ codes are developed to control the system components. A signal of frequency 1.5 MHz, rise time 10.4 ns, fall time 8.4 ns and duty cycle 4.32% is used to trigger the emitter. Scanning measurements corresponding to 40 translations and 40 rotations assuming linear propagation of ultrasound wave are obtained. Total 1600 data files are generated in the whole scanning process. Each data file consists of 7or 8 wave packets (shown in figure 2$a$) and contains 50,002 data points. Each wave packet consist of the transmission data as shown in figure 2$b$.

Distortion in the ultrasound signal may occur in the amplitude and/or in phase due to unequal amplification or attenuation of the various frequency components, changes in the phase relationships between harmonic components of a complex wave. This results in shift of the actual amplitude and formation of a local maxima and minima. The ultrasound signal with amplitude distortion and without any distortion is shown in figure $2d$. The distortion resulted in formation of an additional trough and peak (small amplitude), this affects the detection of the actual peak and trough positions as shown in figure $2c$ and $2d$. Here, red marker shows the detected peaks and troughs. The methods are developed to minimize the effect of such distortion in extraction of the projection data.

Some transmission data have only noise and does not contribute to the actual transmission dataset. Discarding those signal files using threshold filtering based on the Standard deviation. Find threshold value to discard noisy transmission dataset. Noisy data has standard deviation in $10^{-5}$ whereas transmission data has standard deviation in $10^{-4}$. So, the standard deviation threshold ($\tau$) value chosen in this paper is $8.6734 \times 10^{-5}$. If $\tau$ is greater than $8.6734 \times 10^{-5}$, the dataset is classified as transmission data and if less than given threshold then noisy data.

The methods are developed under two main categories, conventional and AI, described below-

## 2.1 Conventional Methods-

i) *Gradient method:* This method is developed on the basis of physical insight of the ultrasound signal. The method developed without involvement of any signal transformation technique to maintain the integrity of the acquired signal. The data is detrended to remove any slanting and divided into sections depending on the number of wave packets. Each section is processed to find out the peaks and the troughs. The peak with highest amplitude is considered to be the transmission peak. The troughs are synchronised according to arrival time to find out the trough with minimum time from the peak in forward sense. The process is repeated for each data file to generate the projection data. The algorithm of the method is shown in figure 3 under Gradient Search Method section.

ii) *Fast Fourier Transform:* The FFT computes the frequency domain representation of the signal, algorithm of the process is shown in figure 3 in the Transform Methods section. The threshold value is set to remove the noise and the unnecessary small amplitude peaks. The windowing technique is used to isolate the region of interest from the signal. Region of interest corresponds to the wave packets containing transmission peak. Further window optimisation is applied to detect the transmission peak.

iii) *Wavelet transform:* Wavelet transform is used to decompose the signal into the set of time extension/contraction of the basis function (called wavelet). This provide the time and frequency information of the analysed signal with better temporal resolution for higher frequencies. The 6 level Symlets (sym5) wavelet function is used to decompose the ultrasound signal. The Shannon energy calculation with moving average filter are used to extract the transmission peaks to generate the projection data. The algorithm of the process is shown in figure 3.

## 2.2 Artificial Intelligence-

Training of the neural network model and features extraction are important stage for any pattern recognition task. Five data files each consisting of 50002 sample data points are used to train the models. The choice of features are a bit abstract and are not a standard in signal processing as these are determined by observation and feature optimization. In total, eight dynamic and morphological features ($F_i$) are obtained for classification of the AI based methods. These features are the function of $X(n)$ and $Y(n)$ which are defined as-

$X(n)$ - The wave is determined by an array describing the displacement at time n.
$Y(n)$ - The displacement of first extrema which comes after time n.

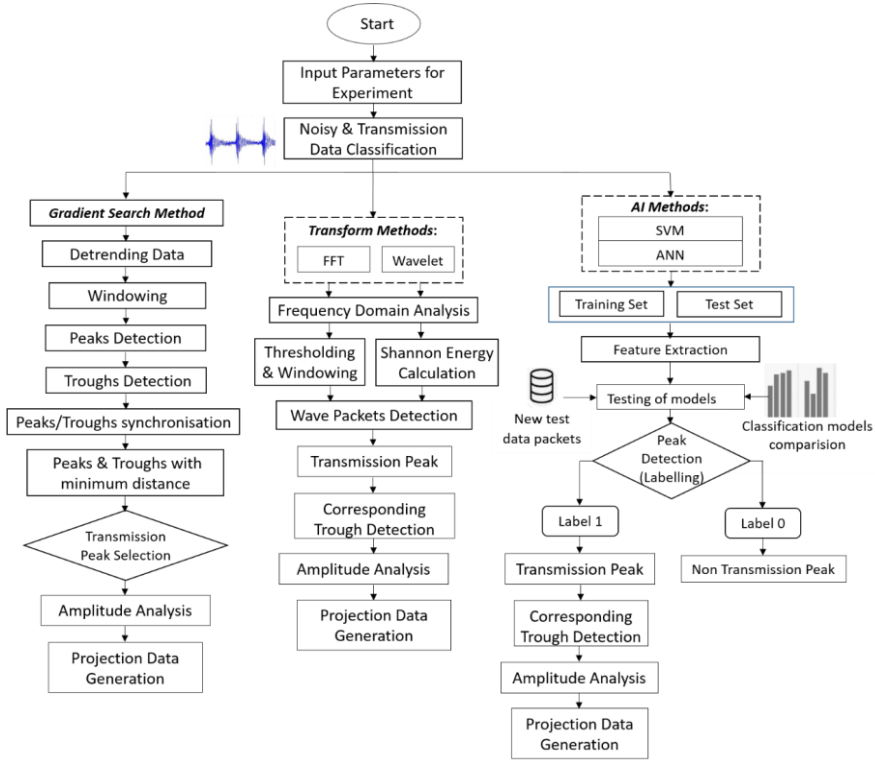

*Figure 3 Methods algorithm.*

The following features were extracted from the signal-

1. From the observation of the training data, most of the peaks classified to be transmission peaks were local maxima or minima so the absolute difference between the next minima or maxima will be maximum. This feature is normalized, as only relative value is required.

$$F_1(i) = \frac{||X(i)| - |Y(i)||}{max(F_1)}$$

2. Average amplitude in the neighbourhood of points. This feature is the ratio of the amplitude of the current point to the average amplitude of $m$ points behind and $m$ points in front i.e. in neighbourhood of the current point.

$$F_2(i) = \frac{|X(i)| * 2 * m}{\sum_{k=i-m}^{i+m} |X(k)|}$$

3. This feature can take only 2 values 0 or 1. This feature is 1 if the current point is local maxima and 0 if not the local maxima.
4. The amplitude relative to the average of the absolute amplitude of all the data points. As the amplitude of the transmission peaks is nearly same, this feature is taken into account as the observed peaks were local maxima. The value of feature is found to be very high for the transmission peaks.

$$F_4(i) = \frac{|X(i)| * n}{\sum_{k=1}^{n} |X(k)|}$$

5. MFCC stands for Mel's frequency cepstral coefficient, these are set of 8-13 cepstral coefficients. For our dataset only the first coefficient was observable, the rest were evaluated to zero; this may be attributed to the high frequency of ultrasound.

6. The ratio of current amplitude to absolute amplitude of the next local minima or maxima. This feature is chosen, as by observation, the peaks were all local maxima and the next peak (trough) was local minima

$$F_6(i) = \frac{X(i)}{|Y(i)|}$$

7. The ratio of current amplitude to maximum amplitude in the neighborhood of the current point. This feature was observed to be 1 for most of the peaks and less than one for other points.

$$F_7(i) = \frac{X(i)}{max([X(i-m)...X(i+m)])}$$

8. The ratio of the current value of feature 1 to the average of feature 1 in the neighbourhood of the current point. This feature was observed to be high for most of the peaks and less for other points.

$$F_8(i) = \frac{F_1(i) * 2 * m}{\sum_{k=i-m}^{i+m} F_1(k)}$$

In machine learning and statistics, classification is a supervised learning approach in which the computer program learns from the input data and then uses this learning to classify new observations. This data set may simply be bi-class (like identifying whether the person is male or female or that the mail is spam or non-spam), or it may be multi-class. Some practical examples of classification problems are speech recognition, handwriting recognition, biometric identification, document classification, etc.. For our task ANN and SVM classification algorithms are used[22, 23]. The problem is formulated in the following manner. All the n points $X(n)$ can be classified in two classes: 1) Non transmission Peak 2) Transmission Peak. The Training examples are vectors in a multi-dimensional feature space, each with a class label. The training phase of the algorithm consists only storing the feature vectors and class labels of the training samples.

The artificial neural network (ANN) classifier uses a set of connected framework of nodes (called artificial neurons) to make predictions about the class of the test sample. Python 3.7.9 and Keras 2.4.3 libraries are used to construct ANN model. Activation function used in the model is Sigmoid and optimizer used is adam. Training is performed in batch size of 100 and epochs of 50. The weight are updated and loss is calculated using categorical cross entropy loss. Cost-sensitive-based learning method is used in this paper to overcome the class imbalance problem. The 0.004 cost is penalized for majority class and very high cost 1 is penalized for minority class.

$$\text{Cost Weight} = \begin{bmatrix} 0 & 0.004 \\ 1 & 1 \end{bmatrix}$$

SVM classifier use complex features to classify the input data into two classes. The values are classified by the decision function. Optimisation of algorithm is accomplished by applying the weight and bias to the classification problems for cost function minimisation. SVM model is built using Scikit-learn 0.24 inbuilt library. Similar steps as the ANN method are followed. In SVM model, same cost is assigned to class 0 and 1. The algorithm of the ANN and SVM methods is shown in figure 3 under artificial intelligence section. Five-fold cross validation is performed on 20 % of training dataset to train and validate the models.

## 3. Results and Discussion

The methods are developed to process the signal acquired from the 2D UCT to extract the projection data. The data is generated for a composite cylindrical specimen of diameter 2.4 cm made up of steel grade 80(outer cylinder) with aluminium grade 1050(inner cylinder). Total 1600 files are generated in the scanning process for a 40X40 data. Only 10 data files are used for training the AI based models, as each file contain 50002 data points and training with more data files result in increased computational time. The digital phantom of the specimen is used to quantify the error in reconstruction of the specimen profile. The projection data extracted from the methods is used for the profile reconstruction. The projection data is shown in the form of sinogram in figure 4 $a)$ to $e)$ for all methods and for phantom in figure $4f)$. The comparative performance of the methods is tested for RMSE, accuracy, F1-score, recall, precision, specificity, Matthews's correlation coefficient (MCC).

The RMSE in the reconstruction results is shown in figure 4 $g)$, it ranges from 0.0189 to 0.0199 and minimum for the ANN based method. Comparative performance of the methods on the basis of accuracy, F1-score, recall, precision, accuracy and Matthews Correlation Coefficient (MCC) in the form of boxplot is shown in figure 5. The green line shows the median values on the boxplot. Each point in boxplot denote results corresponding to one data file (Total 1600). It shows that the conventional signal processing techniques performed better in extracting the amplitude when compared to AI techniques. ANN and SVM data shows the higher deviation in accuracy, F1-score, recall, precision, specificity and MCC. The average F1 score of Gradient based method is 0.9983, FFT based is 0.99990, wavelet based is 0.9926, ANN is 0.8417 and SVM is 0.8523. FFT performed best in terms of the F1 Score. The calculation of F1-Score includes both classes into consideration. It is a better parameter measure because it is not biased toward one class.

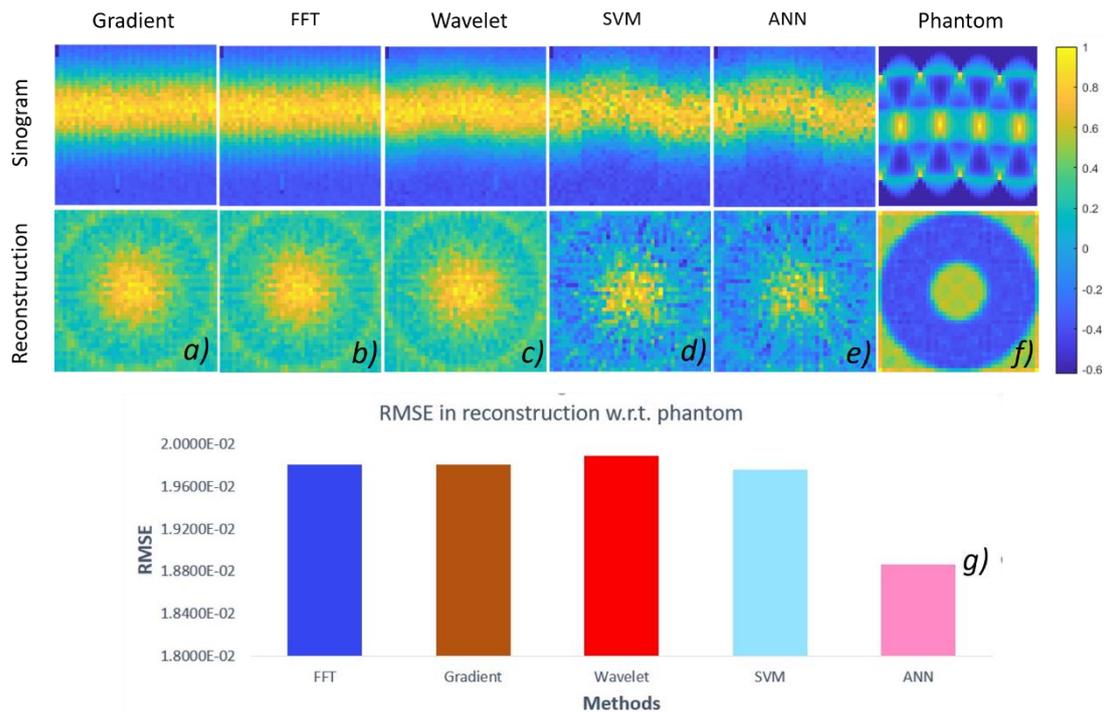

*Figure 4 Quantification of error in reconstruction, a) Gradient, b) FFT, c) Wavelet, d) SVM, e) ANN, f) Phantom, and g) RMSE comparison of all methods.*

Accuracy of all the method is high because its calculation include (true negative (tn) + true positive (tp)) in numerator. Since dataset is highly imbalance in nature, the total negative are very high compared to total positive (true positive << true negative). It resulted in high value of accuracy. This shows that accuracy is not a good measure for imbalanced dataset classification. The average accuracy of all signal and AI methods reported in this paper is 1. Recall of FFT is best out of all methods. Recall gives information about the positive (transmission peak) class of the dataset. Boxplot of wavelet and Gradient based technique include outliers compared to FFT based method. The specificity of models is high because it measures the models ability to identify the negative class (non-transmission peaks). In our application, data is imbalanced, and the majority class is non-transmission peaks. So specificity is high for most of the model. The average precision of SVM is 0.8865, ANN is 0.900, wavelet based method is 0.9874, FFT based method is 0.9998 and Gradient based method is 0.9977. Matthews Correlation Coefficient (MCC) is high only if both classes is classified correctly. It takes all confusion matrix parameters into consideration. FFT based method have least Standard deviation and highest mean out of all methods (Signal Processing as well as AI). The average MCC of Gradient based method is 0.9983, FFT based is 0.9999, wavelet based is 0.9930, ANN based is 0.8730 and SVM based method is 0.8733. The FFT based method found to be performing better in comparison to other methods.

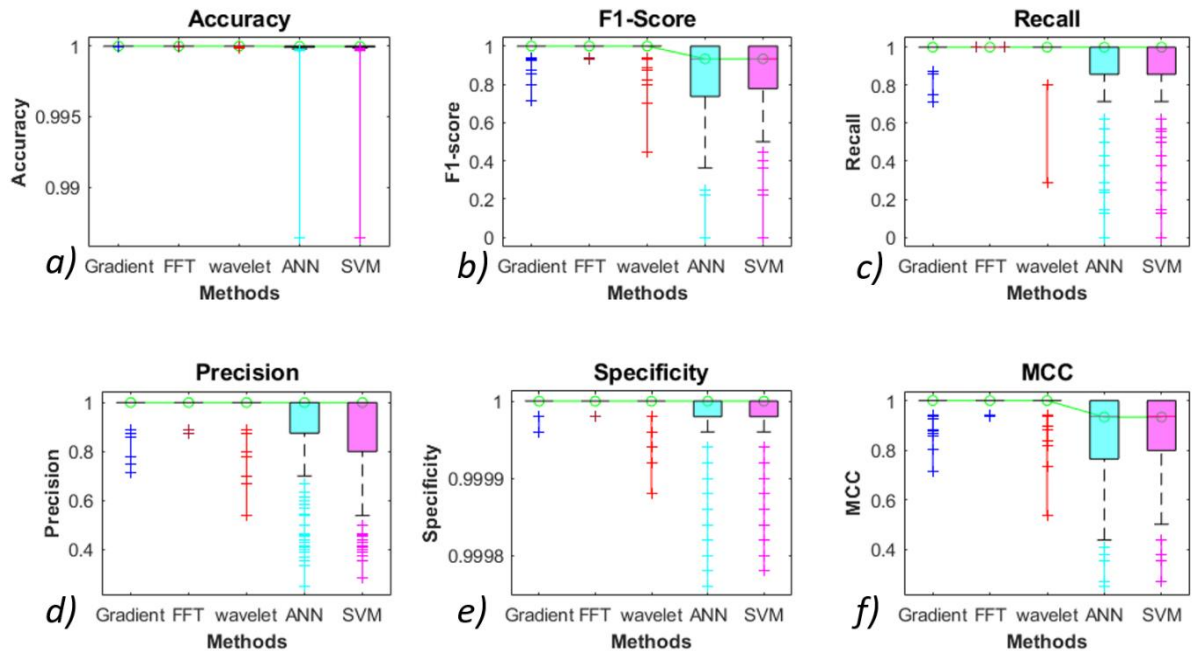

*Figure 5 Methods comparative performance, a) Accuracy, b) F1-scare, c) Recall, d) Precision, e) Specificity, f) MCC*

The RMSE value of the ANN method is minimum while the performance of the ANN method in other performance indices is below average. Also from the view of reconstruction result of the phantom and other methods, the conventional methods showed better performance which is supported by other performance indices. Thus, the RMSE value is not a relative parameter for quantifying the error is the reconstruction. In accuracy, the unexpected high value is because of imbalance nature of the data. In the data used, the maximum value of the true positive can be 8(as each data file contains maximum of eight transmission peaks) then the true negative will be (50002-8), even the value of accuracy in this case comes out to be almost 1. So, it can be regarded as unrelevent parameter for the comparative performance. The developed methods are easy to understand and implement and can be modified for a different data of the similar type.

## 4. Conclusion

The signal processing methods to automatically extract the projection data from the 2D UCT data are presented. The methods correspond to conventional methods: (a) simple gradient based peak detection, (b) Fourier based, (c) wavelet transform and AI based methods: (a) support vector machine (SVM), (b) ANN. Handling a large amount of data and the distortion of the signal due to ultrasound interaction with the specimen are the main challenges encountered in this problem. Another main challenge faced is the selection of feature for the AI based methods.

On the basis of overall performance, FFT based method is performing with least error in recovering the projection data from the UCT signal. In the broad category, conventional methods performance is better in comparison to the AI based methods. The ANN method is found to performing better in case of RMSE error in the reconstruction of the specimen profile. The codes are developed for a general processor and processing is carried out in a sequential manner. Thus, it takes more computational time and processing power. The future work will be to develop parallel processing based codes to improve the computational performance.

## CRediT authorship contribution statement

**Ankur Kumar**: Methodology, Investigation, data measurement, Writing, non-AI methods and Analysis. **Prasunika Khare**: AI methods and analysis, **Mayank Goswami**: Methodology, Investigation, Writing, Visualization, Supervision, Funding acquisition.

## Acknowledgment:

MG acknowledge the Science and Engineering Research Board (SERB), Government of India, for providing support with Grant No. SER-1280-PHY. AK like to acknowledge CSIR Research fellowship. We also acknowledge Mr. Utakarsh Vinayak Parkhi's help in testing preliminary KCNN codes in MATLAB™.